# Maria Goeppert Mayer’s Theoretical Work on Rare-Earth and Transuranic Elements†

Frank Y. Wang
Department of Mathematics
LaGuardia Community College of the City University of New York
31-10 Thomson Avenue, Long Island City, NY 11101
Email: fwang@lagcc.cuny.edu

## Abstract

After the discovery of element 93 neptunium by Edwin McMillan and Philip H. Abelson in 1941, Maria Goeppert Mayer applied the Thomas-Fermi model to calculate the electronic configuration of heavy elements and predicted the occurrence of a second rare-earth series in the vicinity of elements 91 or 92 extending to the transuranic elements. Mayer was motivated by Enrico Fermi, who was at the time contemplating military uses of nuclear energy. Historical development of nuclear science research leading to Mayer’s publication is outlined. Mayer’s method is introduced with the aid of a computer, which enables students to visualize her description of eigenfunctions, particularly the sudden change of spatial distribution and eigenenergy at the beginning of the rare-earth series. The impact of Mayer’s work on the periodic table is discussed.

---

† Submitted to the *Journal of Chemical Education*

**Maria Goeppert Mayer's Theoretical Work on Rare-Earth and Transuranic Elements**

Frank Y. Wang
Department of Mathematics
LaGuardia Community College of the City University of New York
31-10 Thomson Avenue, Long Island City, NY 11101
Email: fwang@lagcc.cuny.edu

Maria Goeppert Mayer was the second of the only two women to date to receive the Nobel Prize in Physics (the first being Marie Curie). Although she is best known among physicists for her nuclear shell model, she also made important contributions to physical chemistry. After the discovery of element 93 neptunium by Edwin McMillan and Philip H. Abelson in 1941,[1] she undertook a quantum mechanical calculation and predicted the occurrence of a second rare-earth series in the vicinity of elements 91 or 92 extending to the transuranic elements. Her work was published in the *Physical Review* on August 1, 1941.[2] In spite of the oversimplications of her method using a statistical model known as the Thomas-Fermi potential, her prediction turned out to be remarkably accurate. This paper guides the reader to use a computer, unavailable to Mayer at the time, to analyze the behavior of the effective potential function and numerically solve the Schrodinger equation. Using simple commands of widely available scientific computing packages such as *Maple* or *Mathematica*, a student can verify Mayer's calculation and visualize her description of the solutions to the Schrodinger equation. Additionally, one can apply her method to make quantitative estimates of energy levels of atoms throughout the periodic table, a task first carried out by Richard Latter in 1955.[3] As stated in an earlier paper published in this *Journal*,[4] there is a large gap between simple textbook problems (the infinite square well, simple harmonic oscillator and hydrogen atom) and a quantitative treatment of atoms with more than one electron. Mayer's paper serves as an excellent supplement for students to develop and strengthen quantum concepts. Before discussing her work, background information about Mayer and the status of nuclear physics and chemistry in her time is outlined. Many students have the impression that history of science is a totally progressive, orderly, and logical development of ideas. The situation of nuclear science research in the 1930s defied such a pattern; through a historical retrospect one will further appreciate the significance of Mayer's work. Although one needs to be acquainted with quantum mechanics to comprehend Mayer's work, the next and last sections on history should understandable for readers equipped with basic knowledge of science.

## Historical Background

Maria Goeppert was born in 1906 in Germany.[5] In 1924 she enrolled at the University of Göttingen, with the intention of becoming a mathematician. At the University she was invited by Max Born, one of the founders of quantum mechanics, to join his physics seminar. Under Born's guidance her interest shifted to physics. As a student of Born, who is a theoretical physicist with a strong foundation in mathematics, she was well trained in the mathematical concepts required to understand quantum mechanics. She received her doctorate in 1930 in theoretical physics, and shortly after she went with her American husband Joseph Mayer, who was appointed as a professor of chemistry at Johns Hopkins University, to Baltimore.

In the following years Maria was occupied with two small children. At the height of the Depression in the early 1930s, opportunities for her to have a normal professional appointment were extremely limited. She managed to obtain an office space at the Physics Department of Johns Hopkins, and undertook scientific research under an unfavorable circumstance. Her collaborators included her husband and Karl Herzfeld, and her main area was applying quantum mechanics to chemistry. She was then one of the few people in America who had a strong background in quantum mechanics. Mayer and Herzfeld's student Alfred Sklar published a pioneering work on the electronic structure of benzene,[6] which has become a classic. The Mayers moved to Columbia University in 1939, and it was at Columbia that she first began to come under the influence of Enrico Fermi.

Fermi received the Nobel Prize in Physics in 1938, and after he attended the award ceremony, he proceeded directly from Stockholm to New York to assume a professorship at Columbia University in order to escape Fascist Italy. Fermi was recognized by the Nobel Committee for his discovery of new radioactive elements produced by neutron irradiation. Fermi and co-workers were inspired by the experiment of Irene Curie and Frederic Joliot, who bombarded elements with the alpha particles (two protons and two neutrons, the nucleus of a helium atom) to create artificial radioactivity.[7] I. Curie, Mme Curie's daughter, and Joliot shared the Nobel Prize in Chemistry in 1935. Curie and Joliot were able to transform aluminum to phosphorus which then beta decayed (through positron emission) into silicon, described by these reactions:

$Al^{27} + He^{4} = P^{30} + n$
$P^{30} = Si^{30} + e^{+} + \text{neutrino}$

After the discovery of neutron by James Chadwick, Fermi and co-workers conceived that neutrons would be more suitable projectiles to activate atomic nuclei. This group conducted an intense series of experiments in which they bombarded every element they could obtain with neutrons. Out of 68 elements they investigated, 47 showed detectable activity. In 1934, they irradiated uranium with neutrons, and found a number of radioactive products. Chemical tests showed that some radioactive products could not be identified with known elements in the neighborhood of uranium. Fermi thought he had discovered transuranic elements.[8] After his uranium experiments were repeated by Otto Hahn and Lise Meitner in Berlin, and I. Curie and P. Savitch in Paris, he named elements 93 and 94 Ausenium and Hesperium respectively. Fermi's conclusion was criticized by Ida Noddack (the co-discoverer of the element rhenium), and she offered an alternative hypothesis—nuclei were splitting into two parts to form elements of much lower atomic number.[9] Noddack's suggestion was ignored, though she was eventually proven to be correct.

While Hahn and Meitner were able to reproduce Fermi's result, they were puzzled by the long chain of observed decays. In late 1937, Curie and Savitch reported a radioactive product produced by uranium nuclei and neutrons to be very similar to lanthanum.[10] In the summer of 1938, Meitner was forced to flee Nazi Germany. Shortly after Meitner's exile, Hahn and Strassmann reexamined the lanthanum claim of Curie and Savitch hoping to prove them wrong, and they found out that barium was present when uranium was bombarded by neutrons. Upon learning Hahn and Strassmann's result, Meitner realized that the uranium nucleus has indeed been split; she and O. R. Frisch named such a process as "fission." They estimated that about 200 MeV of energy would be released through this reaction:

$$U^{235} + n = Ba^{140} + Kr^{93} + 3n$$

With the theory of nuclear fission,[11] it became clear that Curie has been unknowingly splitting atoms, and Fermi's claim of element 93 was really a mixture of disintegration of fission products. Curiously enough, the discovery of the first transuranium element, neptunium, was in turn a by-product of studies conducted of the fission process. Meitner and Frisch's paper ended with a statement "It might be mentioned that the body with the half-life 24 min which was chemically identified with uranium is probably really $U^{239}$ and goes over into eka-rhenium which appears inactive but may decay slowly, probably with emission of alpha particles." ("Eka" is an old prefix meaning "beyond.") When the news of the discovery of nuclear fission reached Berkeley in 1939, several experiments were designed to check and extend the announced results. McMillan irradiated uranium with neutrons, and detected a new beta activity of half-life 2.3 days associated with the 24-minute $U^{239}$. Emilio Segre, a former member of Fermi's group and the co-discoverer of element 43 now known as technetium (below manganese and above rhenium in the periodic table), tested the material giving the 2.3 day activity, and he found that it was unlike manganese or rhenium but more like a rare earth element. Since the rare earth elements are prominent among the fission products, Segre published a paper entitled "An Unsuccessful Search for Transuranic Elements,"[12] and missed the discovery of element 93. It has later been elucidated that uranium 238 can absorb a neutron, instead of undergoing fission, and become uranium-239. Uranium-239 then beta decays to neptunium, which beta decays to element 94 plutonium:

$$U^{238} + n \text{ -> } U^{239} \text{ (23.5 min, } \beta^{-}\text{)-> } Np^{239} \text{ (2.3 days, } \beta^{-}\text{)-> } Pu^{239}$$

Plutonium was first identified by Glenn T. Seaborg and his collaborators in 1941, but it was kept a secret until the end of World War II.

At the time Mayer met Fermi in 1939, fission fragment were once thought by Fermi to be transuranic elements, and a true transuranic element was mistaken as a fission fragment by Segre. Both errors were based on the same faulty assumption with a periodic table shown in Figure 1.[13] Uranium was known to have some similarity to tungsten, and element 93 was thought to resemble rhenium as suggested by the periodic table of the time. In hindsight, Segre could have claim the discovery of neptunium had he correctly interpreted his chemical analysis.

## PERIODIC TABLE - BEFORE WORLD WAR II

| | | | | | | | | | | | | | | | | | |
|---|---|---|---|---|---|---|---|---|---|---|---|---|---|---|---|---|---|
| 1 H | | | | | | | | | | | | | | | | | 2 He |
| 3 Li | 4 Be | | | | | | | | | | | 5 B | 6 C | 7 N | 8 O | 9 F | 10 Ne |
| 11 Na | 12 Mg | | | | | | | | | | | 13 Al | 14 Si | 15 P | 16 S | 17 Cl | 18 Ar |
| 19 K | 20 Ca | 21 Sc | 22 Ti | 23 V | 24 Cr | 25 Mn | 26 Fe | 27 Co | 28 Ni | 29 Cu | 30 Zn | 31 Ga | 32 Ge | 33 As | 34 Se | 35 Br | 36 Kr |
| 37 Rb | 38 Sr | 39 Y | 40 Zr | 41 Nb | 42 Mo | (43) | 44 Ru | 45 Rh | 46 Pd | 47 Ag | 48 Cd | 49 In | 50 Sn | 51 Sb | 52 Te | 53 I | 54 Xe |
| 55 Cs | 56 Ba | 57–71 La–Lu | 72 Hf | 73 Ta | 74 W | 75 Re | 76 Os | 77 Ir | 78 Pt | 79 Au | 80 Hg | 81 Tl | 82 Pb | 83 Bi | 84 Po | (85) | 86 Rn |
| (87) | 88 Ra | 89 Ac | 90 Th | 91 Pa | 92 U | (93) | (94) | (95) | (96) | (97) | (98) | (99) | (100) | | | | |
| | | 57 La | 58 Ce | 59 Pr | 60 Nd | (61) | 62 Sm | 63 Eu | 64 Gd | 65 Tb | 66 Dy | 67 Ho | 68 Er | 69 Tm | 70 Yb | 71 Lu | |

XBD9611-05694.TIF

**Figure 1**: Periodic table before World War II. Fermi irradiated U with neutrons and found products which are chemically similar to Mn and Re; he thought he had found element 93. Before Meitner proposed the theory of nuclear fission, she called the speculated elements 93, 94, 95 and 96 eka-rhenium, eka-osmium, eka-iridium and eka-platinum, respectively. Segre was convinced that element 93 would be chemically similar to Re and Mn, and missed the discovery of element 93, which is actually chemically similar to U.

Segre visited Fermi around Christmas of 1940, and they were intrigued by the possibility that plutonium might be used as a nuclear explosive.[14] If such was the case, it could provide an alternative to $U^{235}$ and avoid the necessity of separating isotopes. For this reason Fermi took great interest in the chemical properties of transuranic elements. He suggested Mayer to calculate the electronic structure of heavy elements based on a statistical model he developed in 1927. In the introduction section of Mayer's paper, she stated: "the chemical behavior of element 93, recently discovered by McMillan and Abelson is strikingly similar to that of uranium and has led these authors to the assumption that a second rare-earth group might start at uranium. In this paper this possibility is discussed from the theoretical point of view." Mayer was to investigate whether the 5*f* shell is filled in the transuranic region of elements. The consequence of such filling, as in the case of the rare-earth elements, is that the outer electrons, which largely determine chemical behavior, remain much the same; this electronic configuration will lead to a series of chemically similar elements.

The current periodic table was first proposed by Seaborg.[15] [16] He used to say that by moving 14 elements out of the main body of the periodic table to a location below the lanthanide series, known as the actinide series, the Swedish Academy awarded him a Nobel Prize. (But don't forget the tedious efforts toward synthesizing and isolating these elements under his leadership.) Mayer's paper concludes with this statement: "in the neighborhood of $Z$=92, the theory predicts occurrence of a second rare-earth group; the first filled 5$f$ level should occur at $Z$=91 or 92." Mayer's conclusion was cited by Seaborg in his paper published in *Nucleonics* in 1949;[17] see the last section for additional discussion.

**Mayer's Method**

An atom with more than one electron is a complex system of mutually interacting electrons moving in the field of the nucleus, and drastic simplification must be made in order to carry out theoretical calculations. A commonly adopted approximation is to treat the quantum states of each individual electron in the atom as being the stationary states of the motion of each electron in a self-consistent central field due to the nucleus and to all the other electrons. This framework of approximation is known as a one-electron theory, and the *s*, *p*, *d* and *f* orbital nomenclature derived for the hydrogen atom remains intact. The treatment of the hydrogen atom can be found in many books. The wave function of the three-dimensional Schrodinger equation is written as the product of the product of a radial function $R(r)$ and the spherical harmonic functions. The radial part of the wave function satisfies an ordinary differential equation, and by the substitution $R(r) = P(r)/r$, the differential equation for the determination of the radial function becomes

$$\frac{d^2P(r)}{dr^2} = \frac{8\pi^2 m}{h^2}[U(r) + \frac{h^2}{8\pi^2 m}\frac{l(l+1)}{r^2} - E]P(r) \tag{1}$$

The meaning of the symbols used in this equation should be obvious from their context; the potential energy is $U(r) = -Ze^2/r$ for hydrogen-like ions. The c.g.s. system of units is used in this section to make a comparison with Mayer's paper easier, and in the next section equations will be in dimensionless form for computation. Equation (1) is considered to be a one-dimensional Schrodinger equation with an effective potential energy

$$U_{eff}(r) = U(r) + \frac{h^2}{8\pi^2 m}\frac{l(l+1)}{r^2} \tag{2}$$

Near the center of the atom, the potential acting on an electron is like that of the Coulomb potential due to the nuclear charge $Z\,e$, and far from the center the behavior of the net potential should be like that of Coulomb potential due to a net charge $e$. The potential energy of the central-field approximation is typically assumed to take the following forms:

$$U(r) = -\frac{Ze^2}{r} \quad r \to 0$$

$$U(r) = -\frac{e^2}{r} \quad r \to \infty$$

The potential energy used in Mayer's work was first developed by Thomas and Fermi independently in 1927. They regarded the electrons in an atom to be a completely degenerate Fermi gas; one can find the Thomas-Fermi model in textbooks.[18 19] In Appendix A, a simplified derivation of such a model is offered. In Mayer's paper, she used this potential energy function,

$$U(r) = -\frac{e^2}{r}[1 + (Z-1)\varphi(r/\mu)] \tag{3}$$

$$\mu = \frac{3^{2/3} h^2}{2^{13/3} \pi^{4/3} m e^2} \frac{1}{Z^{1/3}}$$

where the universal function $\varphi(x)$ is defined by the dimensionless Thomas-Fermi equation

$$\frac{d^2\varphi}{dx^2} = \frac{\varphi^{\frac{3}{2}}}{\sqrt{x}} \qquad \varphi(0) = 1, \quad \lim_{r \to \infty} \varphi(r) = 0 \tag{4}$$

This differential equation with the indicated boundary conditions admits no analytic solution. The complete numerical solution, which is a monotonically decreasing function of *x*, has been tabulated by Fermi. (See online materials.)

The value of the Thomas-Fermi potential is its simplicity, and results from this single universal function are much less accurate than those from the Hartree-Fock method of the self-consistent field. By the very nature of the model, the electron density function is a smoothly varying function of *r*, devoid of the peaks that are characteristic of concentration of the electrons in shells. In Mayer's paper, she declared that inaccuracies of a few units in *Z* are to be expected in the calculation based on the statistical model.

**Numerical Procedure**

The universal function in Equation (4) can be found by numerically solving the differential equation by supplying an initial slope $\varphi'(0) = -1.58807102...$ In the present work, an analytic expression obtained by the author from fitting the tabulated numerical values is used:

$$\varphi(x) = (1 + 0.018631x^{1/2} + 1.38885x - 0.50953x^{3/2} + 0.54206x^2 - 0.102467x^{5/2} + 0.0204851x^3)^{-1} \tag{5}$$

Except for the region of small *r* where the Thomas-Fermi model is inapplicable, the maximum error in this fit is less than 0.2 percent.

To solve the Schrodinger equation numerically, it is convenient to write it in dimensionless form; Equation (1) becomes

$$\frac{d^2}{dx^2}P(x) = \left[u_{eff}(x) - \varepsilon\right]P(x) \tag{6}$$

In this equation radial coordinate *x* is measured in Bohr radius $a_0 = h^2/(4\pi^2 me^2)$ and energy $\varepsilon$ in Rydberg energy $e^2/(2a_0)$, 13.6 eV. The effective potential energy is

$$u_{eff}(x) = -\frac{2}{x}\left[1 + (Z-1)\varphi(x/b)\right] + \frac{l(l+1)}{x^2} \tag{7}$$

with a scaling factor $b = [128Z/(9\pi^2)]^{-1/3} = 0.88534/Z^{1/3}$. The solution to Equation (6) is labeled by the principal quantum number *n* and angular quantum number *l*, and they are related by $n = n_r + l + 1$, where $n_r$ is the number of nodes in $P(x)$ excluding the one at the origin.

Many computer programs have a user-friendly differential equation solver, which can solve the equation numerically and graph the results. The syntax of computer algebra commands is designed to be similar to conventional mathematical symbols, so that one can invoke numerical method without programming experience. Instead of elaborating the theory, we utilize the variable step size Runge-Kutta method built into the solvers, which, as opposed to textbook examples, are more sophisticated and robust. (See online materials.)

We employ the `dsolve` (`numeric`) command in *Maple*, or `NDSolve` command in *Mathematica*, to solve Equation (6), a second-order differential equation. We need to specify the initial conditions and the trial energy. It is clear that $P(0) = 0$, though in practice we use a small number instead of 0 to avoid division by zero. The initial slope $P'(0)$ can be chosen arbitrarily, because we will normalize the wave function eventually. According to quantum mechanics, if a particle is to be bound in a potential well, it can do so only if it has a definite energy. By trying different energies $\varepsilon$, one finds such a definite energy, or eigenenergy, and the corresponding eigenfunction; see an example shown in Figure 2.

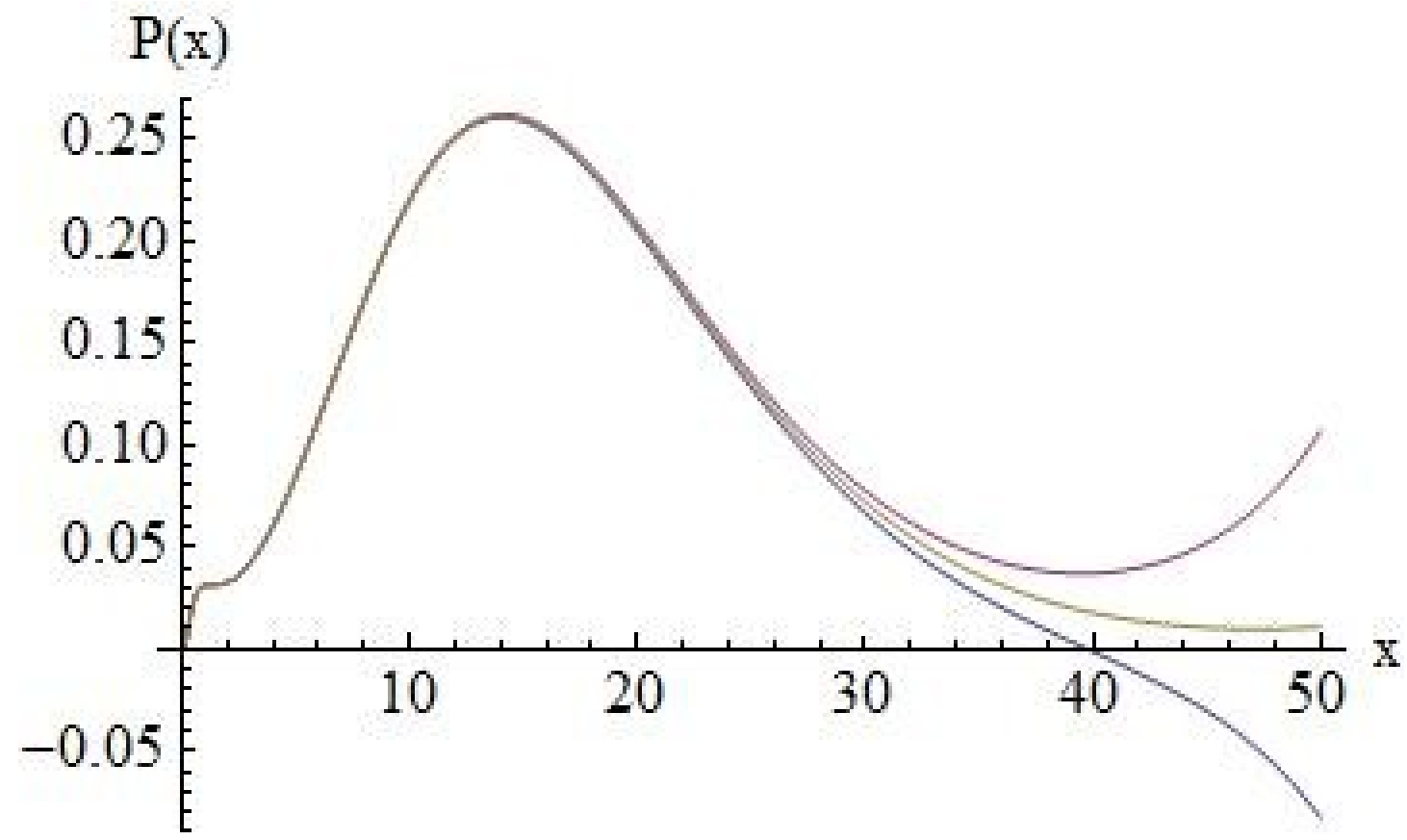

**Figure 2**: Numerical solutions to Equation (6) for $l$=3 and Z=57. Three energies are tried: $\varepsilon$=-0.0668, -0.0669, -0.0670. When the energy is -0.0669 rydbergs, the solution is finite at large distance. We refer to such a solution as the 4*f* orbital according to the definition $n = n_r + l + 1$.

## Mayer's Analysis

With a differential equation solver at hand, students are in a position to verify Mayer's work. In her paper, perturbation corrections such as repulsion among electrons and spin-orbit interaction are not considered (the effect one electron has on another is through the Thomas-Fermi potential); what she called the binding energy is actually a zeroth-order approximation of the one-electron energy in the Thomas-Fermi potential, namely the eigenenergy of Equation (1) or (6).

Mayer recognized that the effective potential in Equation (7) exhibits an asymmetrical double-well structure, separated by a potential barrier, for the angular quantum number $l$=3 only. This two-minimum feature of the potential is the most important realization which would explain the abrupt contraction of the *f* orbitals. Students can graph the effective potential to observe the double-well structure. Mayer used a plot of a function proportional to $r^2 U_{eff}(r)$ to prove that only for $l$=3 does the effective potential energy possess two ranges of negative values separated by a region of positive value; such a plot can be found in online supplement.

In Mayer's paper, she first analyzed the behavior of potential energy function quantitatively; her plots of effective potential energy functions for $l$=3 are reproduced in Figure 3. The valley for larger *r* is very broad and shallow, and the position and depth of the minimum are practically the same as for hydrogen. As *Z* increases, a second minimum of the effective potential develops at small values of *r*. The depth increases very rapidly, and at the same time the valley becomes narrower and the curvature at the minimum increases. A student should try to find the critical value *Z* that a second minimum of the effective potential first appears, and compile a table which lists the positions and depth of minimum values of $u(x)$. Such a task can be accomplished by zooming in the graph, or by applying basic calculus. (See online materials.)

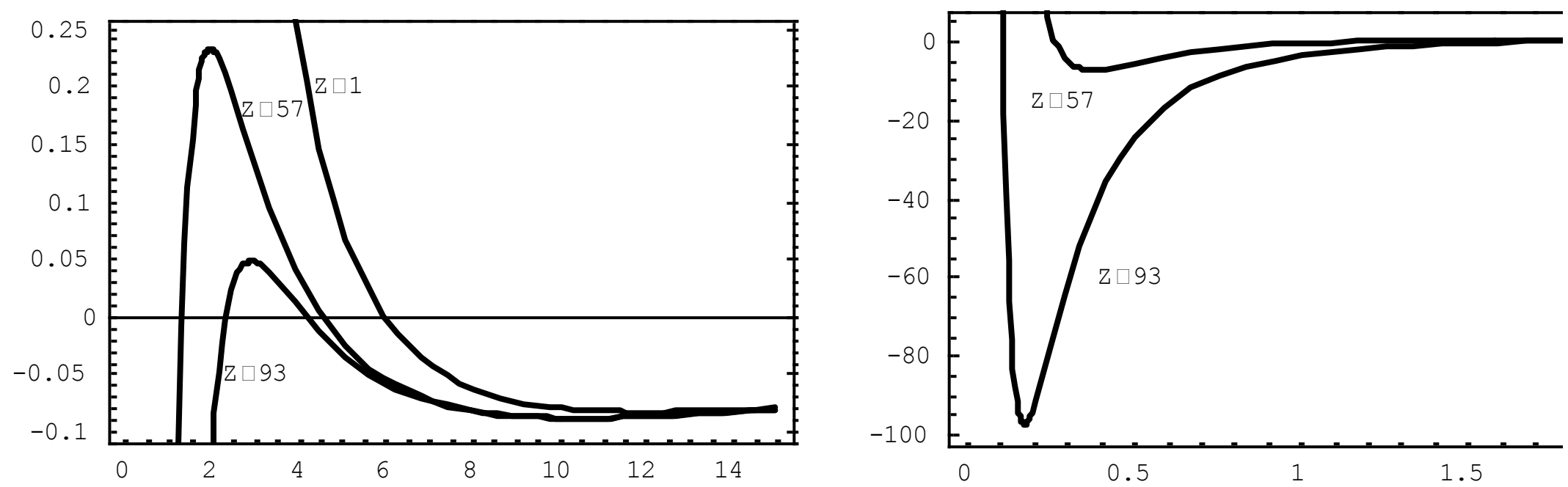

**Figure 3**: Effective potential energy in Equation (7) with $l$=3 and large $Z$ exhibits a double-well structure. Because the inner valley is deep and narrow and can not be conveniently shown with the outer valley in the same figure, the effective potential energy is plotted in two figures for large and small values of distance.

Mayer stated that if the potential barrier between the two valleys were infinitely high, then the two valleys would have independent sets of energy levels. Since the first valley is much narrower, its levels are much more widely spaced than those of the second one. For small $Z$, the first level of the inner valley will be positive. In that case, the lowest level of the total system is that of the outer valley. Since the potential is practically that of hydrogen, and hardly varies with $Z$, the orbital energy of the 4$f$ electrons is approximately constant and about the same as for hydrogen, $1/4^2$ rydbergs. The lowest eigenfunction has two maxima, at the place of the two valleys, but the value of $P$ at the outer maximum will be much larger than the inner one. By examining Figure 2 carefully, one should see a bump at small $x$.

Mayer continued by saying that a value of $Z$ comes when the inner valley is so deep that its first electron level sinks below the first level of the outer valley. At that point the shape of the 4$f$ eigenfunction changes abruptly to one corresponding to an internal orbit. She invoked the fact that experimentally the rare-earth group occurs around $Z$=58. The sudden spatial shrinkage of the 4$f$ orbit agrees with the experimental fact that these functions do not influence the valence properties of the atoms. She performed numerical calculations for $Z$=57, $Z$=60, $Z$=86, $Z$=91 and $Z$=93.

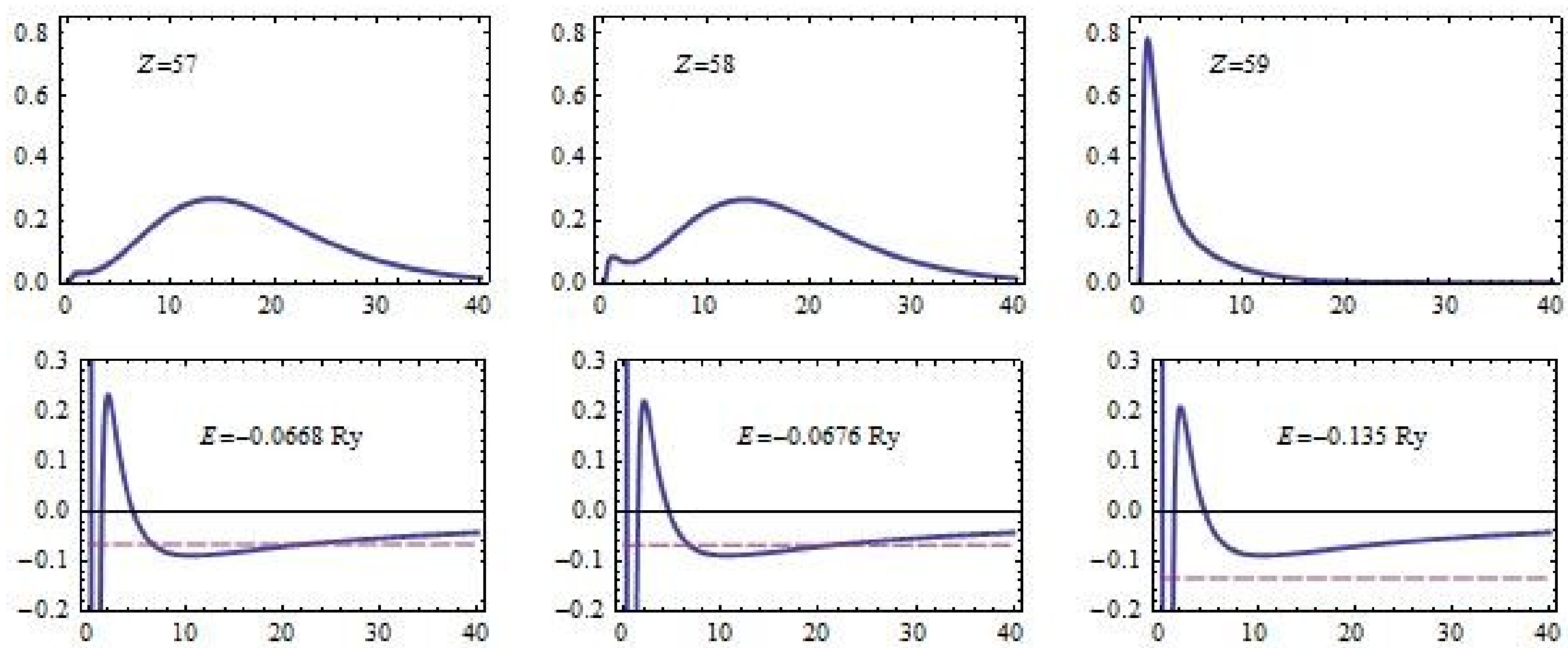

**Figure 4**: The 4*f* eigenfunctions (top row) and effective potential energies with eigenenergies in dashed lines (bottom row) for *Z*=57, 58, and 59. At *Z*=59, the eigenfunction undergoes a radical change: the spatial extension decreases sharply, and the eigenenergy is below the potential energy in the outer valley.

Mayer's description is most easily understood by studying Figure 4. The present calculation shows that at *Z*=58 the eigenenergy is -0.0676 rydbergs, not too different from that of hydrogen, -0.0625 rydbergs. The eigenfunction is located essentially outside the atom. It has two maxima, at *x*=0.90 and *x*=13.8 (the maximum of the 4*f* function of hydrogen occurs at *x*=16; see online materials), but the value of *P* at the second maximum is about three times the value of the first maximum. At *Z*=59, the eigenfunction undergoes a radical change: the energy has dropped to -0.135 rydbergs, which is below the potential energy in the outer valley. The first maximum of *P* occurs at *x*=0.80, but the second one has basically disappeared. The 4*f* function is entirely an inner function.

At the place in the periodic system where the 4*f* function changes its character, the 5*f* function also undergoes changes. For lower values of *Z*, the 5*f* function is hydrogen-like in shape and energy (1/25 rydbergs), with one node somewhere in the outer valley. After the rare-earth elements, the 5*f* function will behave in the outer valley region like a 4*f* hydrogen function. The node of 5*f* function occurs somewhere in the region of the inner valley. As *Z* increases, the energy of the level will remain practically constant around 1/16 rydbergs. We use Figure 5 to illustrate Mayer's point. The 5*f* eigenfunction of hydrogen crosses the *x* axis at *x*=20; this node occurs at *x*=17.8 for *Z*=57. At *Z*=60, the node has come down suddenly to *x*=2.28. For subsequent values of *Z*, up to about *Z*=85, the node moves still further inside, see Figure 6.

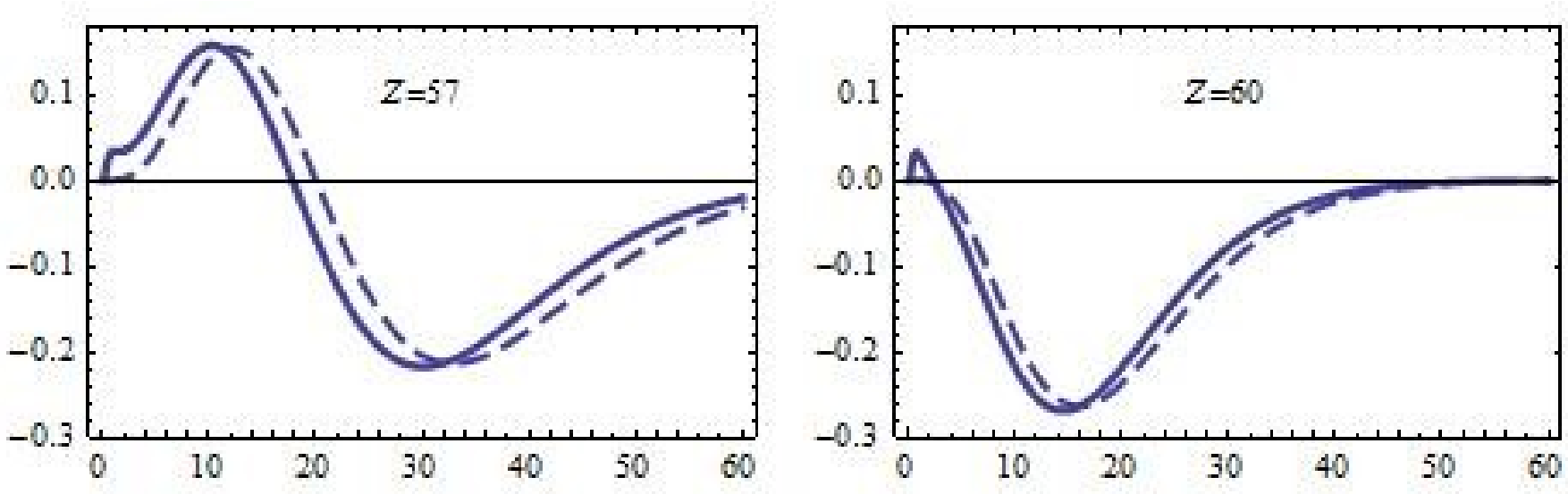

**Figure 5**: As the 4*f* eigenfunctions undergo sudden change near *Z*=58, the 5*f* eigenfunctions change too. For *Z*=57, the 5*f* eigenenergy is -0.0427 rydbergs, and the node of the eigenfunction occurs at *x*=17.8; the hydrogenlike 5*f* orbital (node at *x*=20) is plotted in dashed line. For *Z*=60, the 5*f* eigenenergy is -0.0662 rydbergs, and the node has come down to *x*=2.28; the eigenfunction becomes more like hydrogenlike 4*f* orbital (energy -0.0625 rydbergs), plotted in dashed line.

But then a *Z* value is reached at which the inner potential trough becomes so deep that even the second level of the inner valley drops below the lowest level of the outer valley. One will predict that at that point a second rare-earth group sets in, with all the characteristics of the first one; the energy of the 5*f* level stars to drop, and from then on keeps decreasing with increasing atomic number; the orbits shrink suddenly and become inner eigenfunctions, therefore do not influence the valence character of the atoms.

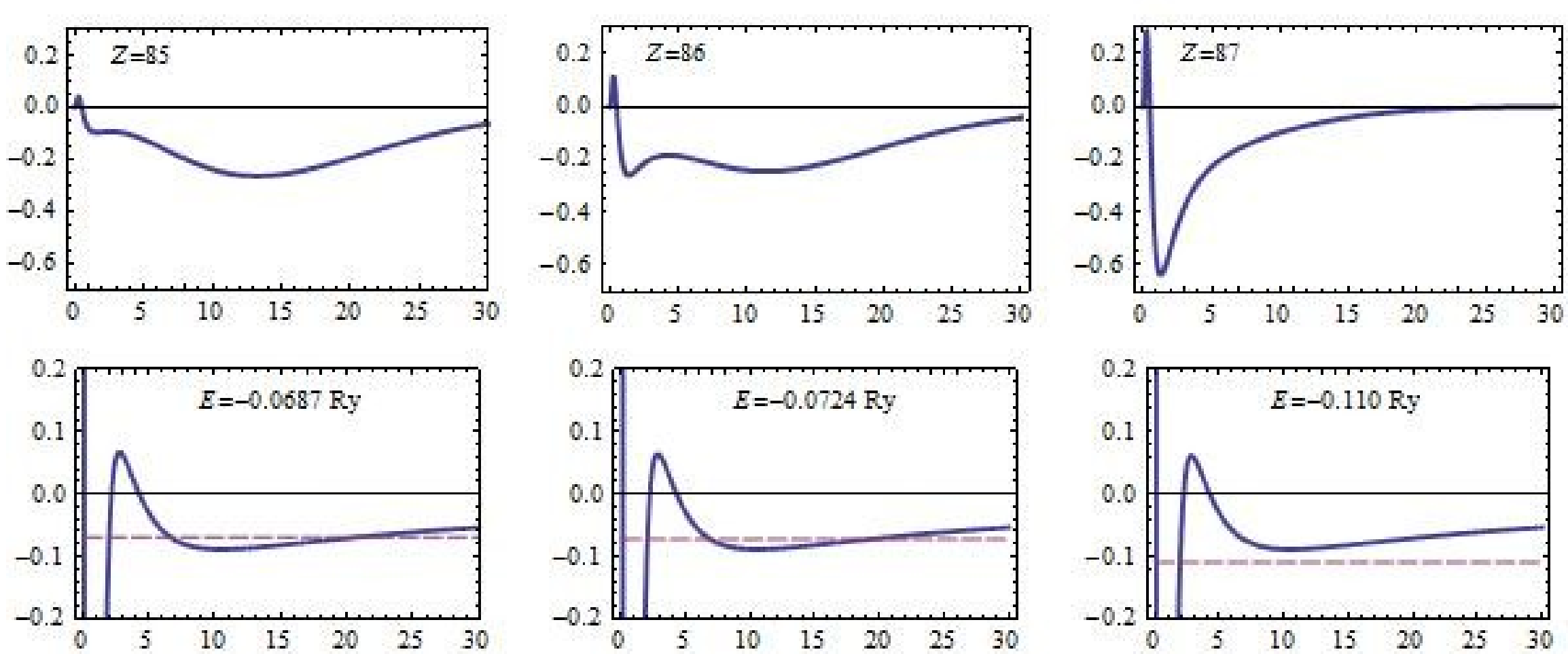

**Figure 6**: The 5*f* eigenfunctions (top row) and effective potential energies with eigenenergies in dashed lines (bottom row) for *Z*=85, 86, and 87. At *Z*=87, the eigenfunction undergoes a radical change: the spatial extension decreases (though the function does not fall off vary rapidly), and the eigenenergy is below the potential energy in the outer valley.

For *Z*=87, the second shrinkage of the 5*f* eigenfunction has already set in. The energy is -0.11 rydbergs, just below the minimum of the outer potential energy. The function *P* has a maximum at *x*=0.256, a node at *x*=0.479, and a minimum at *x*=1.29. The outer minimum has just disappeared, but the function does not fall off very rapidly with distance. In short, the eigenfunction is just on the verge of becoming an inner eigenfunction. At $Z$=91, protactinium, the shrinkage is more complete. (See online materials.) The energy has dropped to -0.556 rydbergs. For $Z$=93, it is -0.893 rydbergs.

In summary, Mayer found that at a critical value of the atomic number $Z$, the *f* orbital starts to be filled. Because the *f* electrons are deep inside an atom, they will no longer contribute to chemical reactions. Mayer concluded that her calculations for the rare-earth series agree moderately well with the experimental facts, with inaccuracies of a few units in $Z$ because of the statistical nature of model. The present calculation predicts that the 4*f* orbital starts to be filled at $Z$=59. Actually, the first 4*f* electron occurs at $Z$=58. Mayer predicted that in the neighborhood of $Z$=92, the first filled 5*f* should occur at $Z$=91 or 92. Experimentally, it occurs at $Z$=91. Mayer's calculation provided a quantum mechanical justification for the remarkable similarity among the rare-earth elements.

## The Aftermath of Mayer's Publication

As mentioned earlier, Mayer's work on rare-earth elements was cited in Seaborg's paper "Place in Periodic System and electronic structure of the heaviest elements" in *Nucleonics* (17) as a theoretical support for his rearrangement of the periodic table. As early as 1922, Niels Bohr envisaged that a second rare earth element might start in element 94. Such an idea had been further discussed by Y. Sugiura and H. C. Urey in 1926, who had displaced the series at $Z$=95 based on the old quantum theory. McMillan and Abelson's experimental evidence was interpreted by them that a new "rare-earth" group of similar elements should start with uranium. If so the chemical properties of elements 95 and 96 should be like those of neptunium and plutonium. But elements 95 and 96 apparently refused to fit this pattern. In 1944, Seaborg suggested the idea that perhaps all the known elements heavier than actinium were misplaced on the periodic table. Seaborg's new concept meant that elements 95 and 96 should have some properties in common with europium and gadolinium. When experiments were designed according to this new concept, elements 95 and 96 were soon discovered.(16) It was unclear whether Mayer's paper might have influenced Seaborg in 1944, but Seaborg in 1949 cited Mayer's paper and said "the latter calculations of Mayer indicate that the energy and spatial extension of the 5*f* eigenfunctions drop sharply at about element 91 and therefore the filling of the 5*f* shell might begin at protactinium or uranium."

The most important extension of Mayer's work is found in the paper by Richard Latter published in 1955.(3) Although the best method of approximate solution of the wave equation for a many-electron atom is the Hartree-Fock self-consistent field calculations, the method is so complex and inconvenient to provide the basis for a systematic discussion of all elements. Without a computer, Mayer was able to calculate the 4*f* and 5*f* orbital energies for selected elements using the Thomas-Fermi model. With a computer Latter implemented a program to calculate the one-electron energy levels through out the periodic table using the Thomas-Fermi and Thomas-Fermi-Dirac models, and compare the results with empirical data. Latter confirmed the behavior for the *f*-series predicted by Mayer, that the 5*f* orbital energy would take on the 4*f* hydrogenic orbital energy; see Figure 5. He also produced curves of the energies of an electron in a central-field orbital (1*s*, 2*s*, 2*p*, etc.) as a function of the atomic number of the neutral atom from 1 to 92. The major features of the periodic system and the sequence of distribution can be understood by Latter's diagram. The essence is that the relative positions of certain orbital energies change with changing $Z$, for instance the 3$d$ curve crosses the 4$p$ curve at $Z$=28. (The actual crossover occurs at $Z$=21.) Some considered Latter's program to be a remarkable achievement—periodicity phenomenon is reproduced by using a single universal function describing a one-electron potential,[20] [21] while some argued that quantum mechanics can only reproduce the periodic system by the use of mathematical approximations, and the Latter diagram is the origin of a misconception of the transition metal electronic configurations.[22] In any case, almost all the physical chemistry or quantum chemistry textbooks merely *describe* the procedure to make quantum calculations for many-electron atoms, thus it is beneficial for students to actually reproduce Mayer's calculation to understand the approximate nature of computational quantum chemistry, particularly the crudeness of the Thomas-Fermi model. In Seaborg's Nobel lecture,[23] he displayed a slide as a pictorial representation of the binding energy in the heaviest elements. We use Mayer's method to reproduce Seaborg's slide and show the crossover of the 5$f$ and 6$d$ energy levels; see Figure 7. The crossover occurs about $Z$=90; at $Z$=91, the 5$f$ orbital energy is below the 6$d$ one. This crude pictorial representation proffers additional support for Seaborg's proposal of the actinide series.

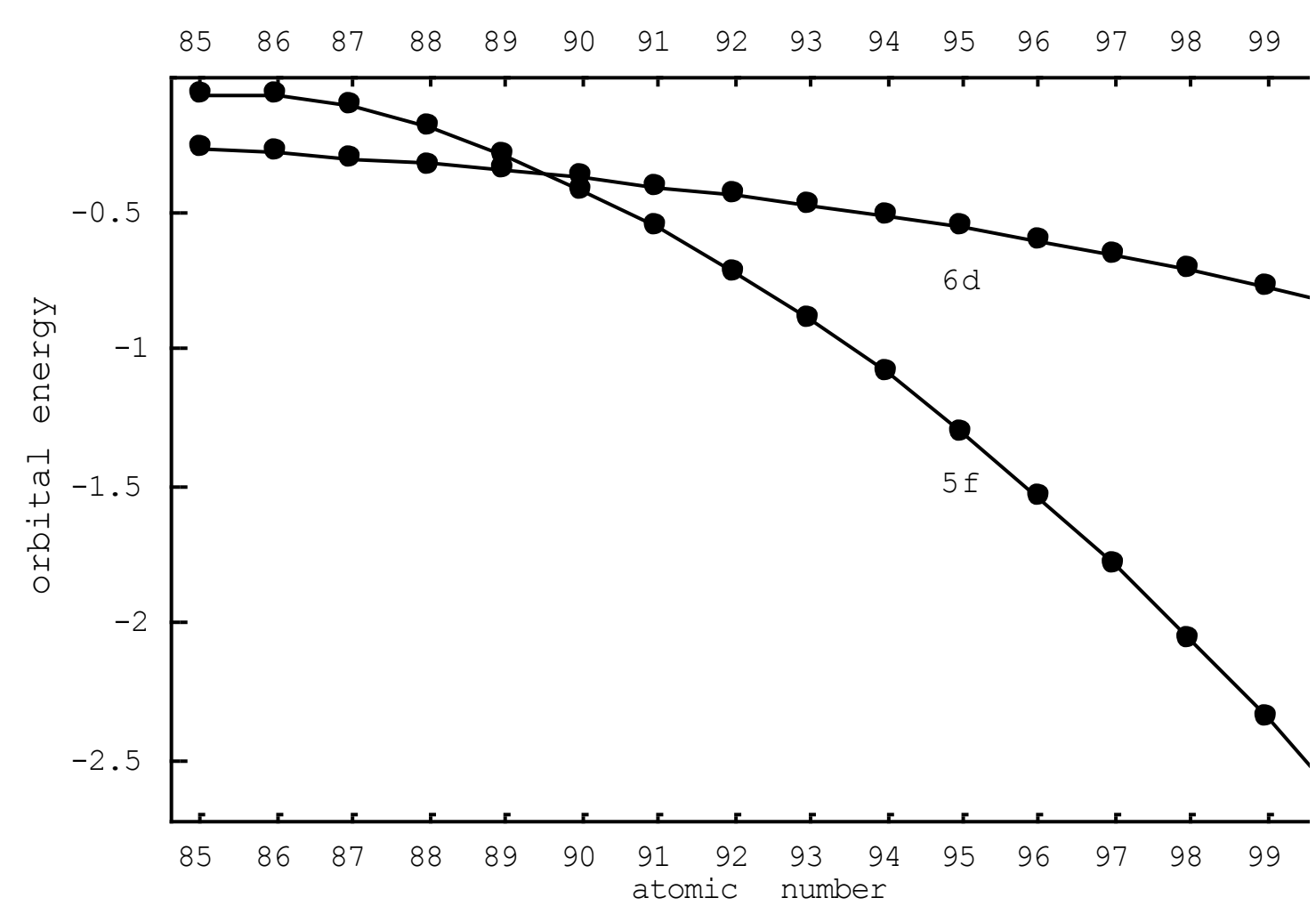

**Figure 7**: Orbital energies of the heaviest elements using Mayer's method. The crossover of the 6$d$ and 5$f$ curves occurs about $Z$=90, and at $Z$=91 the 5$f$ energy level is below the 6$d$ one.

After World War II, the Mayers moved to the University of Chicago where Joseph had been appointed as Professor of Chemistry. At the time, the University's nepotism rules did not permit the hiring of both husband and wife, and Maria's title was "Volunteer Associate Professor of Physics." Later she accepted a regular appointment as Senior Physicist at Argonne National Laboratory, where she made the most important contribution to nuclear physics. She contended that neutrons and protons, in first approximation, move freely throughout the nucleus. After a suggestion by Fermi, she then found that in the nuclei there is a strong coupling of the angular momentum of the orbit and of the spin. Making the assumption of spin-orbit interaction, Mayer explained the phenomenon that there is a special stability associated with certain numbers of protons and neutrons, dubbed as the magic numbers.[24] All these ideas for nuclear physics are basically borrowed from atomic physics.[25] What Mayer's shell model does for nuclei is comparable to what the periodic table does for the chemical elements. For this work she won the Nobel Prize in Physics in 1963. In 1960, she accepted an appointment as a full professor in her own right at the University of California at San Diego, but shortly after her arrival she had a stroke and she had continuing problems with her health. She died in 1972.

## Appendix A: Thomas-Fermi Equation

In applying statistical mechanics to an electron cloud, Thomas and Fermi recognized that it is necessary to use the Fermi-Dirac quantum statistics, based on the Pauli exclusion principle. Let us consider an atom at the absolute zero of temperature: all states with energies below the Fermi energy are occupied and all states with energies above the Fermi energy are unoccupied. If $p_F$ is the maximum value for the electron momentum, the number density $n$ of electrons is

$$n = \frac{2}{h^3}\int_0^{p_F} 4\pi p^2 dp = \frac{2}{h^3}\frac{4\pi}{3}{p_F}^3 \quad \text{(A1)}$$

The factor of 2 is due to spin degeneracy. The charge density is $\rho = -e \cdot n$, and the electrostatic potential energy for an electron is $-e \cdot V$. For an electron confined in a neutral atom, the sum of the kinetic and potential energies is negative, thus the maximum momentum is given by

$$\frac{{p_F}^2}{2m} - eV = 0 \quad \text{(A2)}$$

From Equation (A2) $p_F$ can be written in terms of electric potential, so can the charge density:

$$\rho = -\frac{e}{3\pi^2}\left(\frac{8\pi^2 m}{h^2}\right)^{3/2}(eV)^{3/2} \quad \text{(A3)}$$

The Poisson equation $\nabla^2 V = -4\pi\rho$ relates the electric potential to charge density. Assuming spherical symmetry, the Poisson equation takes the form

$$\frac{1}{r^2}\frac{d}{dr}\left(r^2\frac{dV}{dr}\right) = \frac{4e}{3\pi}\left(\frac{8\pi^2 m}{h^2}\right)^{3/2}(eV)^{3/2} \quad \text{(A4)}$$

Introducing dimensionless variables $w$ and $\varphi$ defined by

$$w = 2\left(\frac{4}{3\pi}\right)^{2/3} Z^{1/3} \frac{4\pi^2 m e^2}{h^2} r$$

and

$$\varphi(w) = \frac{V}{Ze/r}$$

we obtain Equation (4). A number of modifications in the Thomas-Fermi potential have been suggested, and one can find further discussions in Latter's paper.(3) One important cause for a discrepancy between experimental and computed values is the relativistic effects; in Equation (A2), the kinetic energy is of non-relativistic form. According to Latter, the potential used by Mayer gives too large binding energies.

Literature Cited

[1] McMillan, E.; Abelson, P. H. *Phys. Rev.* **1941**, 57, 1185–1186.
[2] Goeppert Mayer, M. *Phys. Rev.* **1941**, 60, 184–187.
[3] Latter, R. *Phys. Rev.* **1955**, 99, 510–519.
[4] Davis, S. L. *J. Chem. Educ.* **2007**, 84, 711–720.
[5] Sachs, R. G. *Biographical Memoirs*, v. 50; National Academy of Sciences: Washington, D.C. 1979; pp 309–328; the biography also appeared in *Physics Today* **1982**, February, 46–51.
[6] Goeppert Mayer, M.; Sklar, A. L. *J. Chem. Phys.* **1938**, 6, 645–652.
[7] Joliot, F.; Curie, I. *Nature* **1934**, 133, 201.
[8] Fermi, E. *Nature* **1934**, 133, 898–899; available at http://dbhs.wvusd.k12.ca.us/webdocs/Chem-History/Fermi-transuranics-1934.html.
[9] Noddack, I. *Angew. Chem.* **1934**, 47, 653; English translation available at http://dbhs.wvusd.k12.ca.us/webdocs/Chem-History/Noddack-1934.html
[10] Seaborg, G. T. *J. Chem. Educ.* **1968**, 45, 278–289.
[11] Meitner, L.; Frisch, O. R. *Nature* **1939**, 143, 239–240; available at http://dbhs.wvusd.k12.ca.us/webdocs/Chem-History/Meitner-Fission-1939.html.
[12] Segre, E. *Phys. Rev.* **1939**, 55, 1104–1105.
[13] Seaborg, G. T. *J. Chem. Educ.* **1989**, 66, 379–384.
[14] Segre, E. *From X-Rays to Quarks*; Freeman: San Francisco, 1980; p 210.
[15] Seaborg, G. T. *Chem. Eng. News*, **1945**, 23, 2190–2193; reprint available in *Modern Alchemy: Selected Papers of Glenn T. Seaborg*; World Scientific: Singapore, 1994; pp 20–23.
[16] Seaborg, G. T. *J. Chem. Educ.* **1985**, 62, 463–467.
[17] Seaborg, G. T. *Neucleonics* **1949**, 5, 16–36; reprint available in *Modern Alchemy: Selected Papers of Glenn T. Seaborg*; World Scientific: Singapore, 1994; pp 149–169.
[18] Landau, L. D.; Lifshitz, E. M. *Quantum Mechanics*, 3rd ed.; Pergamon Press: Oxford, 1977; pp 261–266.
[19] Bethe, H. A.; Jackiw, R. *Intermediate Quantum Mechanics*, 3rd ed.; Westview: Boulder, 1997; pp 83–100.
[20] Pauling, L. *The Nature of the Chemical Bond*, 3rd ed.; Cornell University Press, Ithaca, 1960; pp 55–56.
[21] Ostrovsky, V. N. *Foundations of Chemistry* **2001**, 3, 145–182.
[22] Scerri, E. R. *J. Chem. Educ.* **1989**, 66, 481–483.
[23] Seaborg, G. T. *Nobel Lecture*; available at http://nobelprize.org.
[24] Pauling, L. *General Chemistry*; Dover: New York, 1988; pp 854–856.
[25] Goeppert Mayer, M. *Nobel Lecture*; available at http://nobelprize.org.